\documentclass{article}
\usepackage{mrs2005,epsfig}
\begin{document} 
\title{TRIAXIAL ORBIT-BASED MODELS OF ELLIPTICAL GALAXIES}

\author{Remco van den Bosch\altaffilmark{1}, Glenn van de Ven,
 Ellen Verolme, Michele Cappellari, Tim de Zeeuw}
\affil{Sterrewacht Leiden, Postbus 9513, 2300 RA Leiden,
       The Netherlands.} 
\altaffiltext{1}{email: bosch@strw.leidenuniv.nl}

\begin{abstract} 
  We have developed an orbit-based method for constructing triaxial
  models of elliptical galaxies, which fit their observed surface
  brightness and kinematics. We have tested the method against
  analytical models with general distribution functions. Here, we
  present models that fit integral-field {\tt SAURON}
  observations of NGC~3379 and NGC~4365.
\end{abstract} 

\section{Method} 

We have developed a method to construct triaxial models of elliptical
galaxies, based on an extension of the Schwarzschild (1979) orbit
superposition method. The models fit the observed surface brightness
and (two-dimensional) stellar kinematics. The mass distribution and
gravitational potential are three-dimensional and point-symmetric and
hence include spherical, axisymmetric or triaxial geometries (van den
Bosch et al., in prep). Extensive tests on axisymmetric and triaxial
analytical models show that our method is able to recover general
three-integral distribution functions 
(van de Ven et al., in prep).

\section{Applications}

\begin{figure}  
\begin{center}
\epsfig{figure=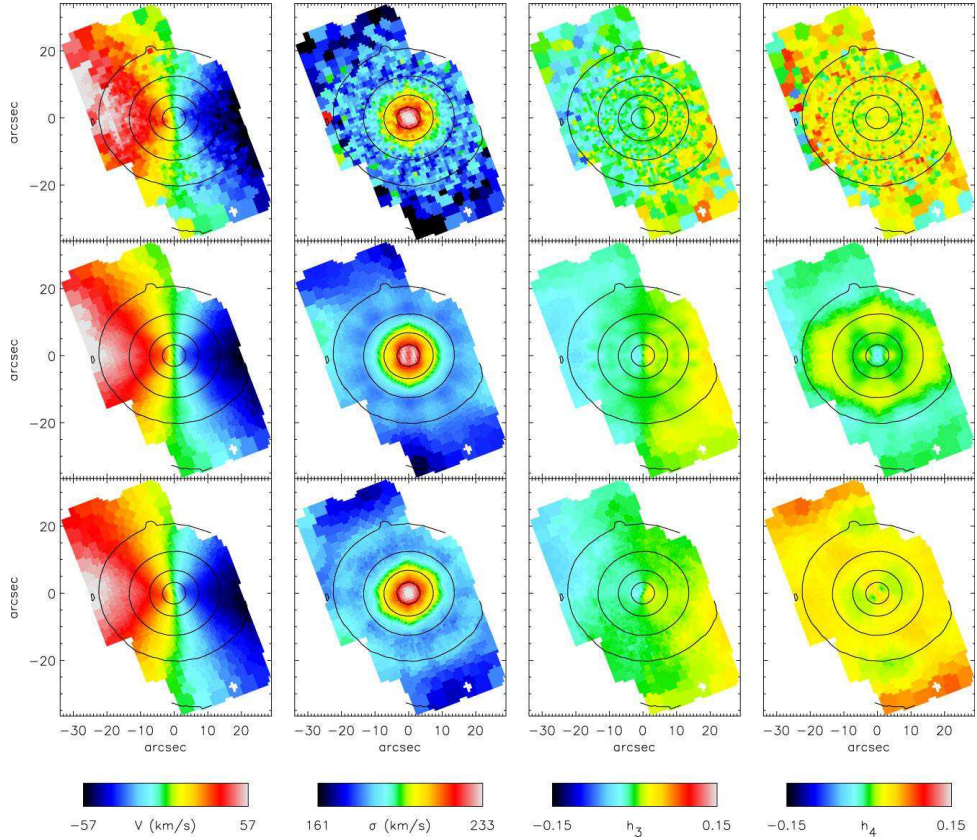,width=12.8cm}  
\end{center}
\caption{From top to bottom: the point-symmetrized {\tt SAURON} stellar kinematics, an axisymmetric
  model and a triaxial model of the elliptical galaxy NGC~3379.  From
  left to right: the mean velocity, velocity dispersion and the
  Gauss-Hermite higher-order velocity moments $h_3$ and $h_4$. The
  black contours show the surface brightness.
\label{N3379}} 
\end{figure}

Figure \ref{N3379} shows an application to the elliptical galaxy
NGC~3379. The kinematics (Emsellem et al.\ 2004) of this galaxy are
reproduced using the models with either an axisymmetric or a triaxial
geometry. The axisymmetric model (Cappellari et al.\ 2005) has axis
ratios 1:1:0.8 and a dynamical mass-to-light ratio ($M/L$) of 3.5
$M_\odot/L_{I,\odot}$. The triaxial model has axis ratios 1:0.9:0.5
and an $M/L$ of 3.6 $M_\odot/L_{I,\odot}$. This demonstrates that it
is possible to reproduce the kinematics equally well with a more
general shape.

\begin{figure}  
\begin{center}
  \epsfig{figure=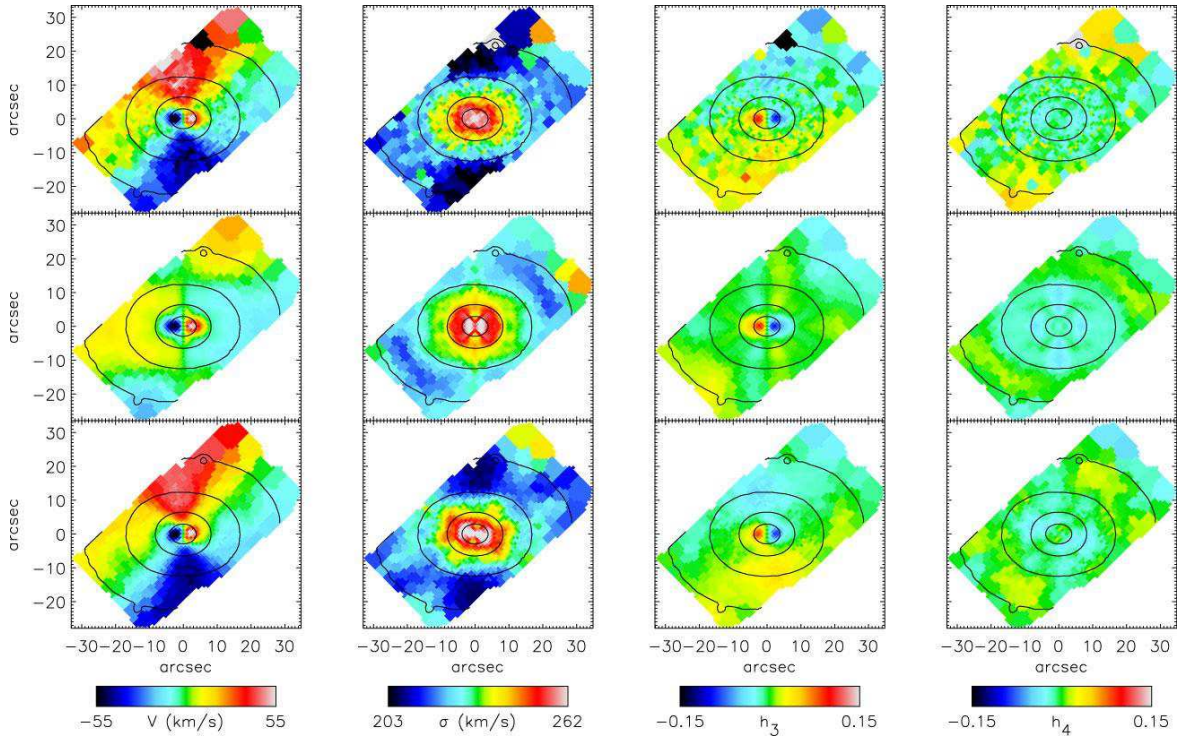,width=15.5cm}
\end{center}
\caption{Similar to Figure~\ref{N3379} for the kinematically decoupled
  core galaxy NGC~4365.
\label{N4365}} 
\end{figure}

Figure \ref{N4365} shows an application to the elliptical galaxy
NGC~4365, with a kinematically decoupled core of which the rotation
axis is $\sim82^\circ$ misaligned with respect to that of the main body
(Davies et al.\ 2001). The axisymmetric model is not able to reproduce
the observed kinematics, while the triaxial model provides a good
fit. The triaxial model has axis ratios of 1:0.7:0.6 and an $M/L$ of
4.0 $M_\odot/L_{I,\odot}$.

The models presented are the first two of a larger sample.
They are based on combined Hubble Space Telescope and ground-based
imaging and stellar kinematics from the {\tt SAURON} survey (de Zeeuw
et al.\ 2002), obtained with the integral-field spectrograph {\tt
  SAURON} (Bacon et al.\ 2001). Such triaxial Schwarzschild models are essential to study the
dynamical $M/L$, super massive black holes, intrinsic shapes, internal
dynamical structure and anisotropy of giant elliptical galaxies.\\
\\
\acknowledgements{
RvdB thanks the Leids Kerkhoven-Bosscha Fonds for travel support.
} 

%
%

%
%

\vfill 
\end{document}